\documentclass{svjour3}                     
\smartqed  
\usepackage{graphicx}
\usepackage{natbib}
\journalname{Earth, Moon and Planets}
\begin{document}

\title{The chemical diversity of comets\thanks{Presented at the workshop
\textit{Future Ground-based Solar System Research: Synergies with Space Probes
and Space Telescopes}, Portoferraio, Isola d'Elba, Livorno (Italy), 8--12
September 2008} }

\subtitle{Synergies between space exploration and ground-based radio
observations}

\titlerunning{Chemical diversity of comets}        

\author{Jacques Crovisier 
\and Nicolas Biver
\and Dominique Bockel\'ee-Morvan
\and J\'er\'emie Boissier
\and Pierre Colom
\and Dariusz C. Lis
}

\authorrunning{J. Crovisier et al.} 

\institute{J. Crovisier, N. Biver, D. Bockel\'ee-Morvan, P. Colom 
\at
LESIA, Observatoire de Paris, 5 place Jules Janssen, 92195 Meudon, France\\
\email{jacques.crovisier@obspm.fr}
\and
J. Boissier 
\at
IRAM, 300 rue de la Piscine, 38406 Saint Martin d'H\`eres, France
\and
D.C. Lis
\at
California Institute of Technology, MC 320--47, Pasadena, CA 91125, USA
}

\date{Received: date / Accepted: date}

\maketitle

\begin{abstract}
    
\sloppy A fundamental question in cometary science is whether the different
dynamical classes of comets have different chemical compositions, which would
reflect different initial conditions.  From the ground or Earth orbit, radio and
infrared spectroscopic observations of a now significant sample of comets indeed
reveal deep differences in the relative abundances of cometary ices.  However,
no obvious correlation with dynamical classes is found.  Further results come,
or are expected, from space exploration.  Such investigations, by nature limited
to a small number of objects, are unfortunately focussed on short-period comets
(mainly Jupiter-family).  But these in situ studies provide ``ground truth'' for
remote sensing.  We discuss the chemical differences in comets from our database
of spectroscopic radio observations, which has been recently enriched by several
Jupiter-family and Halley-type comets.


\end{abstract}

\section{Introduction}
\label{intro}

Among the fundamental questions on comets --- their physical and chemical
nature, their formation and evolution, their relation with the other Solar
System bodies --- is the issue of the possible relationship between their
chemical composition and their orbits.  The dynamical classes of comets point to
various reservoirs.  If these reservoirs are associated with different sites of
formation, one would expect a diversity in the chemical composition of comets,
due to different initial conditions.

\section{Available observations}
\label{observations}

\begin{figure}[h]
\begin{center}
\includegraphics[height=0.80\hsize, angle=270]{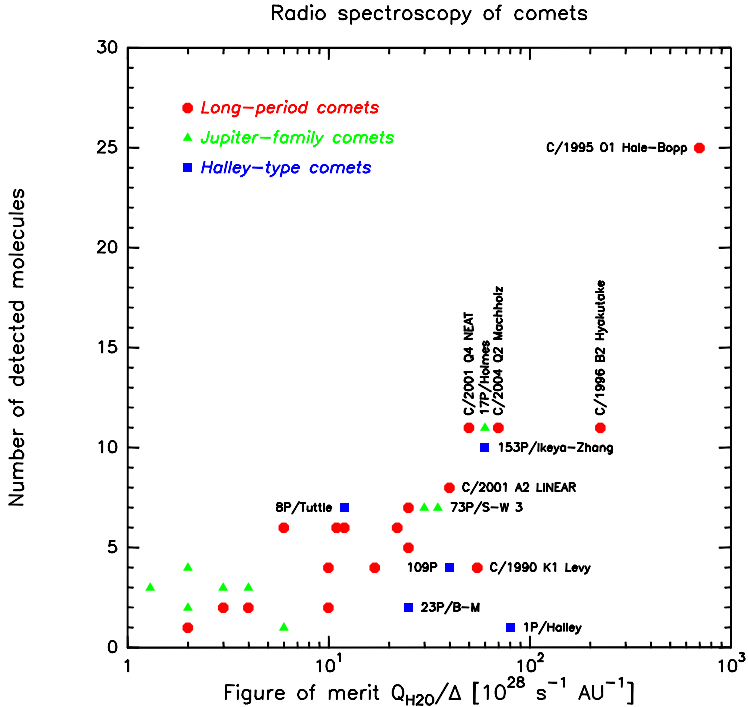}
\end{center}
\caption{Number of molecules, radicals and ions (excepted OH, H$_2$O and 
isotopologues) detected by radio spectroscopy as a function of the
\textit{figure of merit} $Q$[H$_2$O]$/\Delta$ (where $Q$[H$_2$O] is the water
production rate in units of $10^{28}$molecules s$^{-1}$ and $\Delta$ is the
distance to the observer in AU).  This figure of merit, which is roughly
proportional to the observed signal, is convenient for comparing Earth-based
radio observations for different comets.  Only a selection of comets are
identified by names to avoid overburdening the figure.}
\label{fig:1}
\end{figure}

Radio spectroscopy is a major tool for investigating the chemical composition of
comets.  We now have observations on more than 35 comets obtained with the IRAM,
CSO, JCMT and SEST radio telescopes (Fig.~\ref{fig:1}).  \sloppy HCN, CH$_3$OH,
H$_2$CO, CO, HNC, CH$_3$CN, H$_2$S and CS are molecules observed in more than 10
comets and suitable for statistical studies.  Two dozens of rare species and
isotopes are observed only in a few comets (especially in C/1995 O1
(Hale-Bopp)).  The inventory of cometary molecules and their relative abundances
were reviewed by \citet{bock+:2005}.

The water production rate of comets provides a standard for monitoring cometary
activity and a reference for relative abundances in nuclear ices.  Indirect
measurements can be performed from the ground with the 18-cm lines of OH.
About 100 comets (40 comets since 2000) were thus observed with the
Nan\c{c}ay radio telescope \citep{crov+:2002,crov+:2008-ACM}.  Direct
measurements are possible from space with the 557 GHz line of H$_2$O. This line
was observed in several comets by SWAS \citep{neuf+:2000} and 12 comets by the
Odin satellite in 2001--2006 \citep{leca+:2003, bive+:2007-PASS}.  Further
studies of water lines are expected with the Herschel Space Observatory, to be
launched in 2009 \citep[Lellouch et al., this workshop]{crov:2005}.

The conditions of study are quite different for radio observations and for the
space exploration of comets \citep{crov:2007,crov+:2009}.  On the one hand,
Jupiter-family comets are weak and difficult to observe from the ground (with
the exception of comets in undergoing outburst and comets coming close to the
Earth).  On the other hand, space exploration is limited to short-period,
ecliptic comets (a notable exception is 1P/Halley, which was explored at the
price of a large flyby velocity).  This introduces observational bias with
respect to the different families of comets

\section{Recent case studies}

\subsection{Short-period comets}

Our sample of comets was recently enriched by several Jupiter-family comets.
9P/Tempel 1 was the topic of a deep search for molecules as part of the
international campaign of observations in support to the \textit{Deep Impact}
mission \citep{bive+:2007-Icarus}.  73P/Schwassmann-Wachmann 3, a fragmented
comet, was observed in Spring 2006 when it came at only $\Delta \approx 0.08$~AU
to the Earth; its two main fragments B and C were found to have identical
composition \citep{bive+:2008-ACM-73P, lis+:2008}.  17P/Holmes was observed in
October--November 2007 following its huge outburst; the isotopic ratios of C, N
and S could be investigated \citep{bive+:2008-ACM-17P, bock+:2008}.  In
addition, 8P/Tuttle, a Halley-family comet that came close to the Earth, was
observed in December 2007--January 2008 \citep{bive+:2008-ACM-8P}.

\subsection{Comets observed close to the Sun}

Comets at small solar elongations are difficult or even impossible to observe in
the visible/IR from ground-based telescopes or with spacecraft (except
coronagraphs).  Space observatories have a typical constraint $60^\circ < $
solar elongation $ < 120^\circ$.  This is a strong penalty for cometary
observations since comets are brighter close to the Sun (a heliocentric distance
$r_h < 0.20$ AU corresponds to a solar elongation $ < 11.5^\circ$).

However, observations at small solar elongations are possible with some radio
telescopes (e.g., Nan\c{c}ay radio telescope, IRAM 30-m telescope, ALMA).  They
allow us to investigate: 1) gas velocity and temperature; 2) specific mechanisms
such as photolysis shielding; 3) high-temperature sublimation regimes, and to
search for refractories.  Recently, observations at IRAM allowed us to
investigate C/2002 X5 (Kudo-Fujikawa) at $r_h = 0.21$~AU, C/2002 V1 (NEAT) at
0.11~AU and C/2006 P1 (McNaught) at 0.21~AU \citep{bive+:2008-ACM-C/2006P1,
bive+:2009-AA}.

\section{Chemical diversity and taxonomy}

\begin{figure}
\begin{center}
\includegraphics[height=1.0\hsize, angle=270]{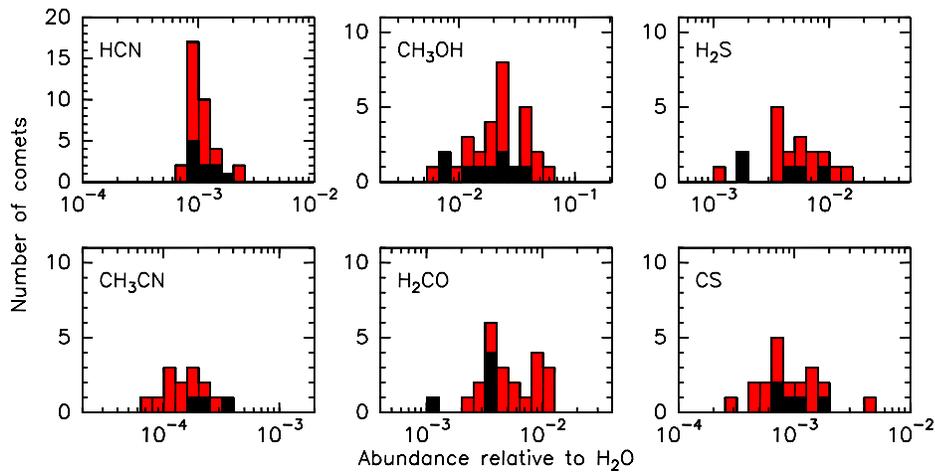}
\end{center}
\caption{Histograms of molecular abundances relative to water from the radio
observations of $\approx 30$ comets.  The black part of the histograms pertains
to Jupiter-family comets, the grey part (red part in the electronic version) to
Oort-cloud comets (long-period and Halley-type).}
\protect\label{fig:2} 
\end{figure}

Our preceding analysis of the chemical diversity of comets from radio studies
was based upon $\approx 20$ comets \citep{bive+:2002-div}.  Our sample is now
extended to more than 35 comets.  Fig.~\ref{fig:2} shows the histograms of the
relative abundances of various molecules.  With the exception of HCN, there are
large variations from comet to comet.  For instance, the distribution of
methanol is spread over a factor of ten.  There is no clear evidence, however,
for the existence of three distinct classes of comets (``normal",``enhanced" or
``depleted" in organics), which were proposed by \citet[and this
workshop]{mumm+:2008} from infrared spectroscopy of a smaller sample of comets.
Fig.~\ref{fig:2} also shows the abundance histograms separately for
Jupiter-family comets and the other comets.  Although the Jupiter-family comets
are under-represented in the sample, there is no indication of a different
distribution for this class of comets \citep{crov:2007, crov+:2009}.

\section{Conclusions}

\begin{itemize}
    
\item The cometary diversity can only be studied by large ground-based or
Earth-orbit observing programmes.  Space exploration accesses only a limited
number of objects among short-period comets.

\item Space exploration provides ``ground truth'' for remote sensing
observations by assessing the link between nucleus ices and coma species.

\item Radio spectroscopy (as well as visible and IR spectroscopy) reveals a
broad chemical diversity among comets.

\item There is no obvious correlation between chemical diversity and the
dynamical classes of comets.

\item Future prospects for radio studies of comets include observations with
increased sensitivity from the ground with ALMA \citep{bock:2008} or from space
with the Herschel Space Telescope \citep[Lellouch et al., this
workshop]{crov:2005} and in situ detailed investigations with Rosetta/MIRO
\citep{gulk+:2007}.

\end{itemize}


\bibliographystyle{crocro_EMP}  

\bibliography{crovisier_elba_emp} 

%
%

\end{document}